**ironbci – Open-source Brain-computer interface with the embedded board to monitor the physiological subject's condition and environmental parameters**

Ildar Rakhmatulin, South Ural State University, Department of Power Plants Networks and Systems, Chelyabinsk City, Russia

ildar.o2010@yandex.ru,

**Abstract**
This manuscript presented brain-computer interface (STM32 and ADS1299) with the embedded board with sensors to monitor the subject's state and environment. To reduce power consumption and device size, we used sensors made in Micro-Electro-Mechanical Systems technology (MEMS) - a gyroscope, accelerometer, and environmental monitoring sensors: $CO_2$, temperature, humidity, ambient sound, and pulse and blood oxygen saturation. Data from the device is transmitted using TCP-PI (UART by Bluetooth) protocol to a computer or mobile device. Open-source https://github.com/Ildaron/ironbci

**1. Introduction**
Electroencephalography (EEG) is a method of measuring brain function, based on the registration of electrical impulses from the electrodes located on a scalp. EEG is used for various purposes, mainly for clinical diagnosis and research, but also for neurofeedback and brain-computer interfaces. Most of the manuscripts in the field of EEG signal processing are devoted to diagnosis of neurological or psychiatric diseases. Kanda et al. [1] used an alpha rhythm to improve the discrimination of mild Alzheimer's. For the same task, Tylová et al. [2] used an unbiased estimation of permutation entropy in EEG analysis. Yu et al. [3] based on the permutation disalignment index researched the functional brain connectivity in Alzheimer's disease.

To register EEG potential differences on the scalp, there are various electrodes: wet or dry, contact, and non-contact. The best signals due to the low impedance are obtained from electrodes that make contact to the scalp using special electrode gel, which have an impedance that usually varies from 200 kOhm before applying the gel and about 5 kOhm after applying the gel.

Today one of the most popular devices for measuring EEG signals in the low-cost segment is the 8-channel Cyton board from the OpenBCI company, but a bare 8-channel board costs about $500 and a 16-channel system including a 3-D printed headset exceeds $ 1,000. Senevirathna et al. [4] measured seven EEG channels with one audio channel and used Bluetooth to transmit data to an external device. However, the board is not compact, the possibility of autonomous use was not considered, and the authors did not present clean EEG signals from the board. Similar disadvantages are with Graham et al. [5], who described in detail the process of developing a compact, low-cost wireless device for measuring EEG signals.

Our paper presents a device that we developed for measuring EEG and environmental signals. The distinctive features are the low cost (open-source), the higher accuracy compared to analogs, and advanced functionality. At the same time, the main points in the creation of the board are described in detail, which will help researchers engaged in similar studies. A promising application is the use as part of the brain-computer interface (BCI). BCIs are used in various fields, one of which is helping people with disabilities through the conversion of EEG signals to a signal that can control various actuators. BCI technology still has several problems that limit its application; one of these is the limited number of controlled functional brain signals and the need for re-calibration of signal processing algorithms during the day. To use neural networks, it is necessary to record EEG for a long time without artifacts. To do this, we need a compact device that can work for a long time in battery-operated mode and with transmission via a wireless connection.

**2. Hardware**
We selected STM32F407VE, which is based on the high-performance 32-bit ARM®Cortex®-M4 RISC core operating at a frequency of up to 168 MHz with 3 SPI ports, 2 I2C, 2 UART. ADC is the main element in devices for measuring EEG signals; the capabilities of BCI devices directly depend on it. There are not many devices on the market specialized in measuring EEG signals. The most popular integrated circuit ADS and their characteristics presented in Table 1.

Table 1. Characteristics of popular ADC models for measuring EEG signals with SPI interfaces

| № | Name | Bitnes s | Data rate, max | Producer | Bias signal | Package | Price, $ | Size, mm | Input impedance, MΩ | Power, mW | Number of inputs |
|---|---|---|---|---|---|---|---|---|---|---|---|
| 1 | ADS1299 | 24 | 16 kSPS | Texas instruments | Yes | TQFP (64) | ≈ 50 | 10x10 | >1000 | 42 | 8 |
| 2 | ADS1191 | 16 | 8 kSPS | Texas instruments | Yes | TQFP (32) | ≈ 15 | 5x5 | >500 | 40< | 2 |
| 3 | AD9653 | 16 | 125 MSPS | Analog devices | Yes | LFCSP (48) | >50 | 7x7 | - | 649 | 8 |
| 4 | AD7770 | 24 | 32 kSPS | Analog devices | No | LFCSP (64) | ≈ 15 | 10x10 | >10 | 136 | 8 |

The characteristics shown in the table are relative and may be interpreted differently depending on the task. The main point when choosing an ADC is the input signal impedance, the ability to measure EEG impedance and control the bias out potential. According to this indicator, the leading role is taken by the ADS1299 from Texas instruments, even though it has been on the market for more than ten years. The main difference between the ADS1299 and the AD7770 it is the presence of an internal multiplexer. The capabilities of this ADC and the characteristics of the multiplexer of ADS1299 analyzed in detail in the manuscripts of Usman et. al [6] and Deepshikha et al. [7].

The need to limit the bandwidth of the signal from the electrode is determined by the fact that the cable capacitance can create a pole in the spectrum with a frequency of up to several kHz. Therefore, it is recommended to limit the bandwidth to at least 1 kHz. But it is worth considering that it is necessary to maintain a bandwidth that exceeds the frequency of the common-mode signal (i.e., 50/60 Hz, 100/120 Hz, etc.). We used a low pass filter as follows, fig.1.

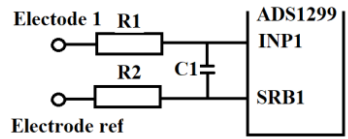

Fig. 1. Low pass filter for electrode input

Texas Instruments recommends a cutoff frequency of 1 kHz to remove high-frequency noise from the signal. Need to add that the use of multi-stage filtering is not possible since it will damage the CMRR of the ADC amplifier. Since ideal multi-stage filtration is extremely difficult to accomplish due to component tolerances, leakage currents, etc.

It is known that the electrode-skin impedances can change over time. These electrode connections must be constantly monitored to check for a suitable connection. The basic idea behind measuring electrode-skin impedance is to inject a known current through the electrode and measure the resulting voltage difference. Since V = I * R, we can easily calculate the impedance. The impedance controlled according to the following scheme, fig. 2 [14]. The resistances R2, R4, R6 in the scheme are close to 0, and the current is 0 in the "electrode-"path.

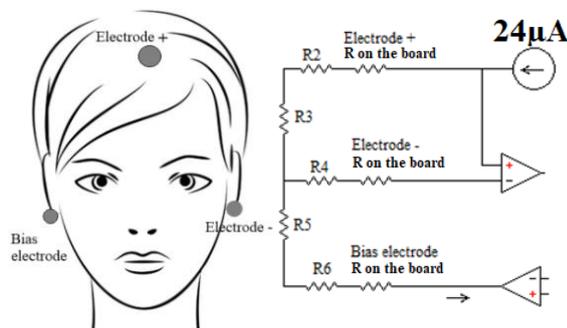

Fig.2. Impedance control process: R2, R4, R6 – Electrode to the skin, R3, R5 – body resistance

An internal excitation signal is used as a current source. This signal is generated by settings that you can control in LOFF: Output Control Register. Such available settings include: Bits [1: 0] FLEAD_OFF [1: 0]: initial frequency of the bit [3: 2] ILEAD_OFF [1: 0]. In our case, the research was carried out at an impedance of 6 kOm.

By means of bias out, we maintain a constant voltage level of the subject body within an acceptable range and minimize any common-mode AC signals in the body. In doing so, BIASout rely on internal connections from the PGA outputs to BIASINV to complete the feedback loop. In this case, the case, interference, and stray characteristics of the cable, as well as the input signal circuit are part of the feedback circuit of the BIAS amplifier, which will allow us to suppress these interferences. The amplifier bias voltage will be centered at 0V (relative value.) Because the amplifier reference voltage is in the middle of the sentence, even in a closed-loop, fig.3.

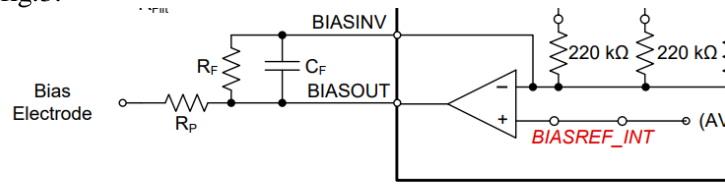

Fig.3. Signal conditioning circuit Bias_out

Biassinv - sets the feedback for adjusting the input voltage to biasout. This mode is established by connecting the input signals from the electrode to BIASINV for the common-mode output. biasref - voltage reference for biasout signal. In our case, we use the internal reference.

In the first, to reduce the effect of external electromagnetic fields the boards with the ADC are placed between two grounded metal plates. It is worth noting that in the low-cost segment, as a rule, engineers miss this component and try to solve the problems with the influence of the magnetic field only due to the bias electrode. We can seal and shape a protective screen, for example, PCB technology company (http://www.pcbtech.ru/) specializes in the production of screens that are made of steel, copper-based alloys, or other materials, with a thickness of 0.1 to 0.5 mm. In our case, completely the protective screen is inconvenient as it takes up additional space. Therefore, we installed two grounded copper alloy plates between the ADC boards.

## 3. Our design

As a reference, we could use the average voltage from all electrodes or use a separate reference electrode. Junghöfer et al [8] in their work point out that, when the number of electrodes is less than 64, it is not advisable to use the averaged reference. Therefore, in our case, a separate electrode was used for the reference, fixed on the earlobe with a clamp. For the manufacture of the board, we used typical fiberglass - FR4 with a nominal thickness of 1.6 mm, lined with 35 μm copper foil of both sides. We used HASL technology as a coating. Board thickness - 1 mm. The developed board consists of three plates with a diameter of 50 mm. 1 - board - power supply with Lippo batteries with capacity 600 ma. The battery power is at the 4.2 V. In the output of the board, an installed step-up transformer, which converts this voltage to 5 V. Thus, from the power board, this power supply in parallel goes to the 2-board with the MCU and the 3-board with the ADS1299. Fig. 4 shows images of blanks for boards and a board with soldered elements.

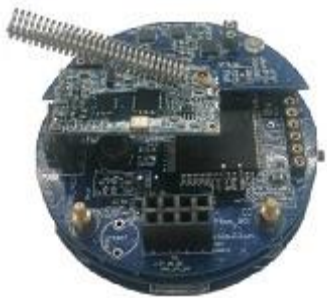

Fig. 4.  General view of the Complete device

For measurement EEG signal we used electrodes from FLORIDA RESEARCH INSTRUMENTS INC (https://www.fri-fl-shop.com/), fig.5.

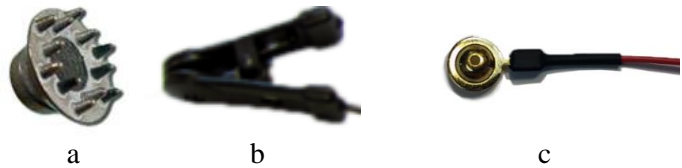

           a                         b                         c

Fig.5. Electrode images: a - Reusable Dry EEG Electrode [TDE-200] or signal measure, b (TDE-430 Silver-Silver Chloride Ear Clip Electrode)- for bias and reference tasks, c - EEG Gold Cup Electrodes

Advisable to use the same electrode metal for all electrodes to prevent the risk of "galvanic" effect.

**4. Device check**
EMG activity due to chewing results in a pronounced and well-studied artifact. Therefore, chewing is often used for the process of checking the correct operation of the EEG device. Allen [10] and Pickworth [11] described in sufficient detail the process of recording the moment of chewing on an EEG encephalogram. Blinking with the eyes also causes an EOG artifact, which is less pronounced. There are many different methods to neutralize this artifact, Sheoran [12] and Borowicz [13] explained how to detect and how to deal with this artifact. In our case, the resulting image of the moment of chewing with the help of the developed device completely coincides with expectations, process measurement, and results in fig.6.

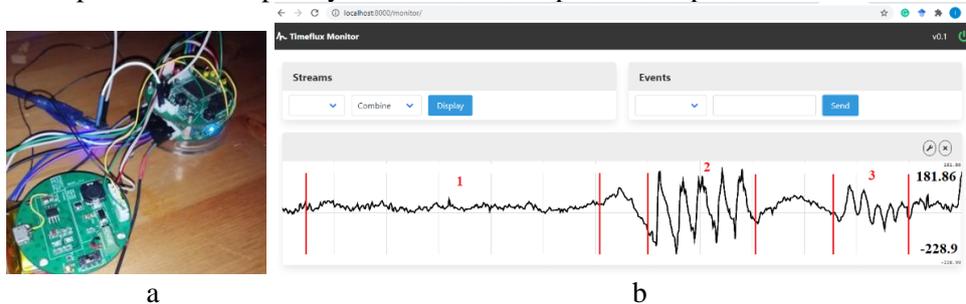

        a                                  b

Fig.6. EEG signal measurement process: a – electronics board, b – dry electrodes, 1 – eyes closed, 2 – chewing, 3 - blinking

Signal processing, we did by Brainflow plugin in Timeflux with pass filter (from 1 Hz to 40 Hz). The noise is about 0.5μV. When the signal is closed with electrodes, a slight network noise of about 0.05 μV is noticeable because of external magnetic fields on the electrical cables. As a result, activity from 8 to 14 Hz showed a typical alpha-wave signal in the occipital lobe of the brain. These results confirm the correctness of the designed device.

**5. Development of a board with sensors**
Microelectromechanical systems (MEMS) are a class of miniature devices and systems manufactured using micro-processing processes. The main criterion for creating MEMS is their size. Usually, it does not exceed 1 mm. MEMS technologies are the forerunner of a relatively more popular field of technology, where device sizes start at 100 nanometers. The term MEMS was originally coined to refer to miniature sensors and actuators operating between the electrical and mechanical areas of a device. Gradually, the term evolved along with the MEMS themselves and encompassed a wide range of different microdevices manufactured using micromechanical processing.

A common feature of electronic systems is the need to exchange information between two or three or ten separate components. There are several standard protocols: UART, USART, SPI, I2C, CAN. Each protocol has its pros and cons, and it is important to know a little about each of them so you can make informed decisions when choosing components or interfaces. We chose I2C since it requires a small number of pins / signals even with many devices on the bus, adapts to the needs of different slaves; and easily supports multiple master devices.

Recognition of EEG signals is primarily associated with identifying the correlation between external conditions. At the same time, if the motor motility is studied well enough, then such factors as temperature, humidity, pressure, noise influence, air quality, the general movement of the object are not fully understood. We have developed an additional board that, if necessary, can be installed on the EEG device. The block diagram of the board with sensors is shown in fig. 7.

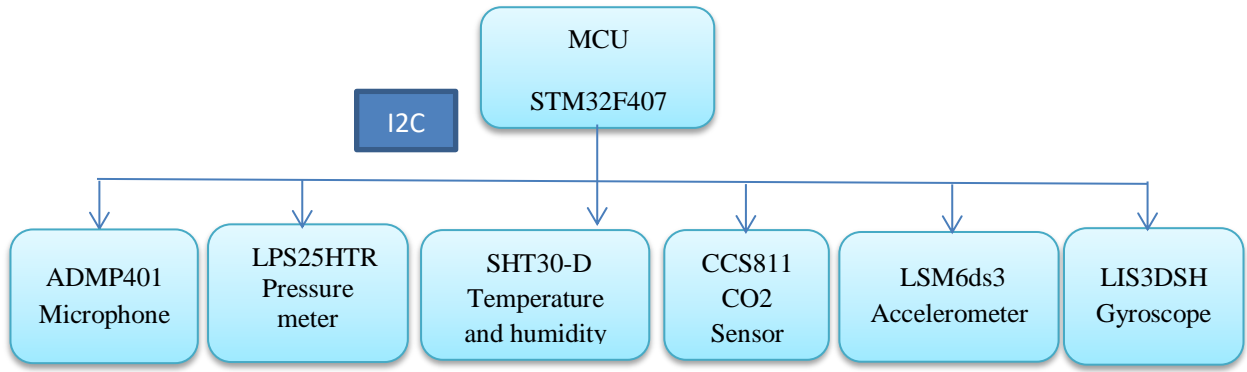

Fig.7. The block diagram of the board with sensors

In the fig.8 showed the PCB boards for sensors, where sensors completed Land grid array packages (LGA).

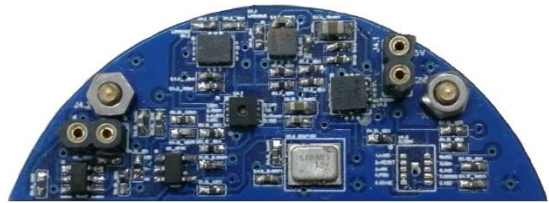

Fig.8. PCB board for sensors

At the first stage, we were unable to find a clear relationship between the sensor signals and the EEG. This is largely because you need a larger number of datasets. It is also necessary to work out the test protocol in more detail. A more complete study of the correlation between the EEG signal readings and the environmental indicators will be carried out in further studies.

**Discuss and conclusion**
This manuscript describes in some detail the process of creating a board based on the ADS1299 ADC. As a result, a device with a peak input noise of 0.5 µV and a common-mode signal rejection ratio of 110 dB in the 0-50 Hz band was developed. The ability to capture from 8 to 24 EEG channels with a sampling rate of 250 to 1000 samples per second was implemented. Apart from the presence of a board with sensors, the main difference of this board from analogs is low power consumption, high accuracy, and compactness, which allows using this board for a long time without recharging.

Presented developed additional board, which includes sensors for monitoring the state of the environment - temperature, humidity, air quality. For physiological activity - gyroscope, accelerometer. To control irritation and speech activity - a microphone sensor. The sensors are made in mems technology, which made it possible to install their data compactly on one board with a diameter of 50 mm. The sensors are connected to the microcontroller using the I2C protocol. These sensors are designed to collect a dataset, which ultimately will allow us to determine the correlations between various external factors and the EEG signal. Ultimately, this information, together with the mathematical filters that are implemented in the software, will allow you to get the purest EEG signal.

In this manuscript, special attention is paid to combating common-mode noise. Suppression of which is carried out, both at the expense of the hardware part and using external shields and shielded cables. A wi-fi wireless connection allows you to transfer data directly to a server that implements software signal filtering. With a 1200mAh LiPo battery, the system can operate for up to 9 hours. The device is designed for continuous use and data collection for the subsequent use of neural networks. Our task is to involve more enthusiasts in the BCI development area due to open sources of the project and the low cost of the hardware. That will allow you to collect datasets and exchange them in an open format.

In the future, it is planned to increase the operating time of the device without recharging due to the use of various modes of the microcontroller's work - Energy saving mode from stm32, such as Low Power Run (LP Run).

With EEG measurement in microvolts, there are many seemingly insignificant moments that ultimately can affect the result of measuring the EEG signal - the choice of the gel, the choice of the method of attachment, various methods of implementation for noise reduction, etc. Each of these moments can contribute difficult to interpret noises into the system and requires additional research.

The device can be used in places with different electromagnetic activity. If the system is exposed to high-frequency electromagnetic interference, it is advisable to add very small common mode capacitors to the inputs to filter high-frequency common-mode signals. If these capacitors are added, then the capacitors should be 10 or 20 times smaller than the differential capacitors to ensure the minimum CMRR impact.

In the future, we will move the filtering work to any host processor with which you communicate, if it has a higher bus bandwidth than the microcontroller. Now, a server is being developed to receive information from the developed device via the tcp-ip protocol.